\documentclass[12pt]{article}

\newcommand{\sll}{/\kern-4pt l}
\newcommand{\slP}{P\kern-7pt/\kern2pt}
\newcommand{\slq}{q\kern-5.5pt/}
\newcommand{\slv}{v\kern-5pt\raise1pt\hbox{$\scriptstyle/$}\kern1pt}
\newcommand{\Li}{{\rm Li}}

\begin{document}

\begin{flushright}
MZ-TH/00-08\\
hep-ph/0003116\\
March 2000
\end{flushright}
\vspace{0.5cm}
\begin{center}
{\Large\bf Theory of Heavy Baryon Decay\footnote{Invited talk given by
J.G.~K\"orner at the $3^{\rm rd}$ International Conference on $B$ Physics and
$CP$ Violation (BCONF99), Taipei, Taiwan, 3--7 Dec 1999}}\\[1truecm]
{\large S. Groote and J.G. K\"orner\\[.5truecm]
Institut f\"ur Physik, Johannes Gutenberg-Universit\"at,\\[.2truecm]
Staudinger Weg 7, D-55099 Mainz, Germany}
\vspace{1truecm}
\end{center}

\begin{abstract}
We discuss various topics in the theory of heavy baryon decays.
Among these are recent applications of the Relativistic Three Quark Model to
semileptonic, nonleptonic, one-pion and one-photon transitions among heavy
baryons, new higher order perturbative results on the correlator of
two heavy baryon currents and on the semi-inclusive decay 
$\Lambda_b \rightarrow X_c + D_s^{(\star)-}$.
\end{abstract}
\vspace{1truecm}
\begin{center}
To be published in the proceedings
\end{center}

\newpage

\section{Introduction}
This review contains four different topics on heavy baryon decays. First we
discuss some recent theoretical determinations of the quasielastic
$\Lambda_b\rightarrow\Lambda_c$ form factor in HQET where there are many
different results in the literature. The quasielastic
$\Lambda_b\rightarrow\Lambda_c$ form factor has also been calculated in the
Relativistic Three Quark Model (RTQM) which is an all-encompassing tool for
the description of exclusive heavy baryon decays. We briefly describe the
RTQM model and discuss various applications of the RTQM. As concerns QCD sum
rules we describe some recent three-loop results on the finite mass baryon
current-current correlator at ${\cal O}(\alpha_s)$ which is a new result
important for QCD sum rule calculations. Finally we discuss the semi-inclusive
decays $\Lambda_b\rightarrow X_c + D_s^{(\star)-}$ where ${\cal O}(\alpha_s)$
and $O(1/m_b^2)$ have been recently calculated. This is an important mode for
$\Lambda_b$-decays with an expected branching ratio of $\approx 10 \%$.

\section{The quasielastic $\Lambda_b \rightarrow \Lambda_c$ form factor}
There exist many different results on the the quasielastic
$\Lambda_b\rightarrow\Lambda_c$ form factor in the literature. The predicted
slope values of the form factor range from $\rho^2=0.33$ to $\rho^2=2.35$. In
the simplest approach one takes a heavy quark -- light diquark model and
describes the transition by a one-loop Feynman diagram
\cite{efimov,ebert,haghighat}. One takes $M_Q=m_Q$ and local point coupling
factors $g_1$ and $g_2$ for the quark-diquark-baryon vertices whose strengths
are fixed by the compositeness condition. The compositeness condition is
nothing but the field theoretic equivalent of the familiar quantum mechanical
concept of wave function normalization. In the heavy quark limit the result
of such a calculation is given by the form factor
$\Phi(\omega)=(\omega^2-1)^{-1/2}\ln(\omega + \sqrt{\omega^2-1})$
which is familiar from the $\omega$-dependent renormalization of the heavy
quark current. An expansion in terms of powers of $(\omega-1)$ shows that the
form factor is correctly normalized at the zero recoil point $\omega=1$, has
a slope of $\rho^2=1/3$ and a convexity of $c=2/15$. The form factor
$\Phi(\omega)$ is rather flat when e.g.\ compared to the heavy meson form
factor where experimentally one finds slope values of $\approx 1$.
$\Phi(\omega)$ lies within the inclusive HQET sum rule bounds derived by
Chiang \cite{chiang} but must nevertheless be discarded since it oversaturates
the semileptonic inclusive rate $\Lambda_b \rightarrow X_c +l^- + \bar{\nu}_l$
as recently shown in \cite{melic}.

Improvements on this simplest approach lead one to the Relativistic Three
Quark Model (RTQM) \cite{ivanov0}. In the improvements one incorporates
binding effects by replacing $M_Q=m_Q$ by $M_Q=m_Q+\bar{\Lambda}$. The vertex
is softened through introduction of a nonlocal vertex and one introduces a
true three-quark structure by replacing the $(ud)$-diquark by single $u$, $d$
quarks.

\section{The Relativistic Three Quark Model}
According to the changes mentioned above the RTQM treats the decay
$\Lambda_b \rightarrow \Lambda_c + W^-_{\rm off-shell}$ in terms of a two-loop
Feynman diagram with nonlocal vertices including binding effects. The result is
that the $\Lambda_b \rightarrow \Lambda_c$ form factor becomes steeper.
Depending on the choice of spin vertex structure for the heavy baryons the
slope increases from the aforementioned $\rho^2=1/3$ to $\rho^2=0.75\div 1.35$
depending on the choice of spin vertex structure to be discussed later on.

The RTQM is an all-encompassing and versatile tool for the description of
heavy baryon decays in terms of a Feynman diagram description. The number of
parameters associated with the nonlocality of the vertices, the binding
effects and the values of the constituent quark masses is reasonably small
and their values lie within common expectations. Many of the parameters are
already fixed from light baryon decays where the RTQM also applies. For the
loop integrations one uses the $\alpha$-parametrization in its exponential
form. This introduces $n$ $\alpha$-parameters $\alpha_1,\ldots,\alpha_n$ for
$n$ propagators and consequently $n$ integrations. One introduces a Laplace
transform to facilitate the vertex form factor integration which is left to
the very end. The exponential $\alpha$-parametrization allows one to do the
tensor loop integrals directly through differentiation, i.e.\ without use of
the Passarino-Veltman expansion. One transforms to spherical type variables
which leaves one with one radial type integration and $(n-1)$ angular type
integrations. All $(n+1)$ numerical integrations including the Laplace
transform can be done with ease. In fact the spherical integrations can also
be done analytically but the ease of the numerical integration does not
warrant this effort. We shall now discuss several applications of the RTQM
to heavy baryon decays.

The dependence of the Isgur-Wise function on the choice of the vertex spin
structure for the heavy baryons was investigated in \cite{ivanov1}. For the
$\Lambda_Q$-type baryons both the effective couplings
\begin{equation}
J_{\Lambda_Q}^1=\bar\psi_Q\psi_{[u}^TC\gamma_5\psi_{d]}\qquad
J_{\Lambda_Q}^2=\psi_Q\psi_{[u}^TC\gamma_5\slv\psi_{d]}
\end{equation}
correctly describe the coupling of the $\Lambda_Q$ to a heavy on-shell quark
and two light off-shell quarks in the limit of Heavy Quark Symmetry (HQS).
When inserted into the relevant two-loop diagram both coupling structures
reproduce the required leading order HQET form factor structure including the
unit normalization at zero recoil. However, one finds that the slope of the
Isgur-Wise function depends on the choice of vertex structure. For the three
choices $J_{\Lambda_Q}^1$, $\frac12(J_{\Lambda_Q}^1+J_{\Lambda_Q}^2)$ and
$J_{\Lambda_Q}^2$ one finds slope values of $\rho^2=1.35$, $1.05$ and $0.75$.
The fact that  $\rho^2(J_{\Lambda_Q}^1)>\rho^2(J_{\Lambda_Q}^2)$ agrees with
the sum rule analysis of \cite{grozyak} although the difference in slope
values in \cite{grozyak} is not as large. Note that
taking the geometric mean of the two currents leads to a constituent type
vertex structure where the projector $(\slv+1)/2$ projects onto the large
components of the light quark fields.

Finite mass effects in heavy $\Lambda_{Q_1}\rightarrow\Lambda_{Q_2}$
transitions were analyzed in \cite{ivanov2} by replacing the heavy quark
propagators in the Feynman diagrams by the full propagator. This was effected
by the replacement
\begin{equation}
\frac{i}{l\cdot v+\bar\Lambda}\frac{\slv+1}2\rightarrow
  \frac{i(\slP+\sll+m_Q)}{(P+l)^2-m_Q^2}
\end{equation}
where $(P+l)$ is the momentum of the heavy quark and $l$ is a loop momentum.
For $\Lambda_b\rightarrow\Lambda_c$ the rate is decreased by $9.3\%$ in
qualitative agreement with the findings of \cite{dosch}. The decrease was
found to be even larger for $\Lambda_c\rightarrow\Lambda_s$ where the strange
quark was treated as a heavy quark in the reference rate for the sake of
comparison. It is clear that an expansion of the full propagator in terms of
powers of $1/m_Q$ would allow one to systematically explore higher order
$1/m_Q$-effects in these transitions as e.g.\ in the zero recoil normalization
of the relevant zero recoil $\Lambda_{Q_1}\rightarrow\Lambda_{Q_2}$ form
factor.

In \cite{ivanov3} the RTQM was used to calculate exclusive nonleptonic decays
of heavy baryons. There are so-called factorizing and nonfactorizing
contributions to these decays. The nonfactorizing contributions had never been
calculated before. In the Feynman diagram approach they involve a genuine
three-loop calculation which was done in \cite{ivanov3}. As a sample result
one finds that in the nonleptonic decays $\Lambda_b\rightarrow\Lambda_c+\pi^-$
the nonfactorizing contributions amount to $-20\%$ and $-28\%$ in the parity
violating and parity conserving amplitudes, repectively, with an ensuing
reduction in rate of $\approx 40\%$. The nonfactorizing contributions are
therefore not negligible. A multitude of exclusive nonleptonic decays have
been calculated within the RTQM model involving $(\bar{s}c)(\bar{u}d)$,
$(\bar{b}c)(\bar{u}d)$ and $(\bar{b}c)(\bar{c}s)$ transitions \cite{ivanov3}.
         
The RTQM model is also well suited for heavy flavour-conserving one-pion and 
one-photon transitions between heavy baryons. The one-pion transitions are
described by two-loop diagrams where the pion couples to a single light quark
line. The $1/f_\pi$ coupling of chiral perturbation theory effectively appears
through the quark level Goldberger-Treiman relation $g_\pi=2m_q/f_\pi$. Many
one-pion transitions have been calculated including transitions from excited
states \cite{ivanov4,ivanov6}. The results are remarkably close to the results
of using the constituent quark model \cite{yan1,yan2,salampion} for the light
quarks even though the light quarks are fully off-shell in the RTQM model.

For one-photon transitions the transverse on-shell photon couples only to the
light quarks in the leading order of the heavy quark expansion with a coupling
strength given by the light quark charge. In addition one has to include
contact graphs to assure gauge invariance of the one-photon transitions. These
are generated according to the path integral formalism of Mandelstam. Again
the results of the RTQM model \cite{ivanov5,ivanov6} are remarkably close to
the constituent quark model calculation \cite{yan3,salamphoton}. The relation
of the RTQM description of one-photon transitions to the chiral approach
remains to be explored, in particular to the recent calculation of \cite{pich}
which contains also chiral loops.

What we have discussed so far are some basic applications of the RTQM in the
heavy baryon sector. Further work is in progress on the decays of double heavy
baryons, on magnetic moments of heavy baryons and on heavy flavour-conserving
nonleptonic charm and bottom baryon decays.

\section{Finite mass baryonic current-current correlator at
  ${\cal O}(\alpha_s)$}
The calculation of the spectral density associated with the baryonic
current-current correlator is important for QCD sum rule applications. We want
to report on some advances we have made in the calculation of the
${\cal O}(\alpha_s)$ radiative corrections to the spectral density with one
finite mass quark mass and two zero quark masses \cite{finite}. The
calculation involves the evaluation of two-scale three-loop Feynman diagrams
which only became possible due to recent technical advances \cite{alex} in
three-loop technology. Taking the appropiate limits we recover previous
results derived for the zero mass case \cite{pivovarov} and for the infinite
mass case \cite{grooteky}.  In the mesonic case the corresponding calculation
has been done some time ago showing that radiative corrections to the spectral
density can become quite important \cite{broadhurst}.

The basic object of study is the vacuum expectation value of the time-ordered
product of two baryonic currents $<TJ(x)J(0)>$. In spinor space its Fourier
transform is expanded along the spinor matrix structures $\slq$ and $m$ with
coefficients $\pi_q(q^2)$ and $\pi_m(q^2)$. We concentrate on the invariant
$\pi_m(q^2)$ which has associated with it a spectral density $\rho_m(s)$ for
which we shall present two- and three-loop results. Using the simplest
possible current $J=\Psi(u^TCd)$ and writing
\begin{equation}
\rho_m(s)=\frac1{128\pi^4}s^2\left\{\rho_0(s)\left(1+\frac{\alpha_s}\pi
  \ln\left(\frac{\mu^2}{m^2}\right)\right)+\frac{\alpha_s}\pi\rho_1(s)\right\}
\end{equation}
we obtain the Born term two-loop contribution
$\rho_0(q^2)=1+9z-9z^2-z^3+6z(1+z)\ln z$ where $z=m^2/q^2$. In the
$\overline{\mbox{MS}}$ scheme
the radiative
three-loop contribution is given by
\begin{eqnarray}
\lefteqn{\rho_1(s)\ =\ 9+\frac{665}9z-\frac{665}9z^2-9z^3
  -\left(\frac{58}9+42z-42z^2-\frac{58}9z^3\right)\ln(1-z)}\nonumber\\&&
  +\left(2+\frac{154}3z-\frac{22}3z^2-\frac{58}9z^3\right)\ln z
  +4\left(\frac13+3z-3z^2-\frac13z^3\right)\ln(1-z)\ln z\nonumber\\&&
  +12z\left(2+3z+\frac19z^2\right)\left(\frac12\ln^2z-\zeta(2)\right)
  +4\left(\frac23+12z+3z^2-\frac13z^3\right)\Li_2(z)\nonumber\\&&
  +24z(1+z)\left(\Li_3(z)-\zeta(3)-\frac13\Li_2(z)\ln z\right)
\end{eqnarray}
Writing $q^2=(m+E)^2$ we can perform a threshold expansion of the spectral
density in terms of powers of $E/m$. We write the leading order result in a
factorized form in order to facilitate comparison with HQET. One has  
\begin{eqnarray}
m\rho_m&\buildrel{m\rightarrow\infty}\over=&\frac1{128\pi^4}E^5
  \left\{1+\frac{\alpha_s}\pi\left(\frac12\ln\left(\frac{m^2}{\mu^2}\right)
  -\frac23\right)\right\}^2\ \times\nonumber\\&&\times\
  \left\{1+\frac{\alpha_s}\pi\left(4\ln\left(\frac\mu{2E}\right)
  +\frac2{45}\left(10\pi^2+273\right)\right)\right\}.
\end{eqnarray}
The first bracket is the square of the appropiate HQET matching coefficient
$C(m/\mu,\alpha_s)$ first derived in \cite{grozyak} and the second bracket is
the appropiate result for the leading order HQET spectral density
$\rho^{\rm HQET}(E,\mu)$ first derived in \cite{grooteky}. We have checked
that the zero mass limit of the general spectral density reproduces the result
of \cite{pivovarov}. Work is in progress on the momentum spectral density
$\rho_q(s)$ and on correlators of baryonic currents with arbitrary spin
structure and on sum rule applications of the spectral densities.

\section{The semi-inclusive decay $\Lambda_b \rightarrow X_c + D_s^{(*)-}$}

Following the analysis of the semi-inclusive B-meson decays
$\bar{B}\rightarrow X_c+D_s^{(*)-}$ in \cite{aleksan,mauser1} we looked at the
corresponding semi-inclusive $\Lambda_b$-decays. The $\Lambda_b$ decays are
potentially more interesting because of the possibility to observe $\Lambda_b$
polarization effects in this decay. At the leading order of $\alpha_s$ and the
heavy quark mass expansion one expects branching ratios of
$BR(\Lambda_b\rightarrow X_c+D_s^-)=3.2\%$ and 
$BR(\Lambda_b\rightarrow X_c+D_s^{*-})=(4.4(L)+2.4(T))\%$ where we have
separately listed the longitudinal ($L$) and transverse component ($T$) of the
spin 1 $D_s^{*-}$. The two components can be separately measured by an angular
analysis of the subsequent decays $D_s^{*-}\rightarrow D_s^-+\gamma$ and
$D_s^{*-}\rightarrow D_s^-+\pi^0$. The same holds true for the measurement of
polarization effects \cite{mauser2} which will not be discussed here.

In \cite{mauser2} we calculated the perturbative ${\cal O}(\alpha_s)$ and the
nonperturbative corrections to these decays using the factorization hypothesis.
Numerically one finds:
\begin{eqnarray}
\Lambda_b\rightarrow X_c+D_s^-\ :\ \quad
  \hat\Gamma_S&=&\phantom{0.65}(1-0.096-0.013)\nonumber\\
\Lambda_b\rightarrow X_c+D_s^{*-}\ :\quad
  \hat\Gamma_L&=&0.65(1-0.110-0.034)\nonumber\\
  \hat\Gamma_T&=&0.35(1-0.108+0.026)\nonumber\\
  \hat\Gamma_{L+T}&=&\phantom{0.35}(1-0.096-0.009).
\end{eqnarray}
In order to clearly exhibit the percentage changes the rates have been
normalized to their respective Born term rates. The second and third
figures in the round brackets of Eq.(6) refer to the perturbative
${\cal O}(\alpha_s)$ corrections and the nonperturbative kinetic energy
correction, respectively. The perturbative corrections are negative and quite
uniform. They amount to $\approx 10\%$. The nonperturbative corrections range
from 0.9\% to 3.4\% with differing signs. The longitudinal mode dominates the
rate into $D_s^{*-}$'s. The $L/T$ rate ratio $\Gamma_L/\Gamma_T$ decreases by
6.8\% from 1.86 to 1.73 after applying the perturbative and nonperturbative
corrections.

The corresponding semi-inclusive $b \rightarrow u$ decays 
$\Lambda_b \rightarrow X_u + D_s^{(*)-}$ are suppressed due to the smallness
of $V_{bu}$. They are nevertheless of interest for the analysis of so-called
wrong sign $D_s^{(*)-}$'s \cite{aleksan}. Numerically one finds \cite{mauser2}
\begin{eqnarray}
\Lambda_b\rightarrow X_u+D_s^-\ :\ \quad
  \hat\Gamma_S&=&\phantom{0.73}(1-0.169-0..013)\nonumber\\
\Lambda_b\rightarrow X_u+D_s^{*-}\ :\quad
  \hat\Gamma_L&=&0.73(1-0.178-0.029)\nonumber\\
  \hat\Gamma_T&=&0.27(1-0.115+0.030)\nonumber\\
  \hat\Gamma_{L+T}&=&\phantom{0.27}(1-0.161-0.001).
\end{eqnarray}
The dominance of the longitudinal mode in the decay
$\Lambda_b \rightarrow X_u + D_s^{*-}$ is now more pronounced. Also the 
radiative corrections are no longer uniform leading to a substantial 13.3\%
change in the $L/T$ rate ratio due to the perturbative and nonperturbative
corrections. It would be interesting to study these semi-inclusive $\Lambda_b$
decay modes including the $L/T$ composition of the $D_s^{*-}$'s at future
colliders.

\end{document}